# Executive Accountability Systems and the Environmental Violations of State-Owned Enterprises in China

Lihua Liu[1], Yi Chen[1], Mingli Xu[2*]

1. School of Accounting, Guangdong University of Foreign Studies, Guangzhou 510006, China
2. School of Politics and Public Administration, South China Normal University, Guangzhou 510006, China

*Corresponding author: Mingli Xu

E-mail addresses: lihualiu75@gdufs.edu.cn (L. Liu), 20241510003@mail.gdufs.edu.cn (Y. Chen), minglixu@m.scnu.edu.cn (M. Xu)

**Acknowledgements**

We thank Gang Qiao and Yue Chen and seminar participants at Guangdong University of Foreign Studies for helpful suggestions. We acknowledge the financial support from the National Natural Science Foundation of China (Grant No. 72202076), Guangdong Planning Office of Philosophy and Social Science (Grant No. GD25YSG03), Guangdong Basic and Applied Basic Research Foundation (Grant No. 2024A1515012683), Guangzhou Science and Technology Bureau (Grant No. 2025A04J3449). All errors are our own.

# Executive Accountability Systems and the Environmental Violations of State-Owned Enterprises in China

**Abstract:** Executive accountability is increasingly viewed as a critical mechanism for improving corporate environmental performance, especially in state-owned enterprises (SOEs) that dominate high-emission sectors such as energy, infrastructure, and heavy industry. This study examines whether China's Accountability System for Irregular Operations and Investments (ASIOI) curbs environmental violations in SOEs. Exploiting the staggered regional implementation of ASIOI as a quasi-natural experiment, we find that the policy leads to a significant reduction in SOE environmental misconduct. Drawing on a criminology-based cost-benefit framework, we identify three underlying mechanisms: strengthened internal controls, increased green investment, and enhanced green innovation. Further analyses reveal that the deterrent effect of ASIOI is more pronounced in SOEs that exhibit weaker regulatory oversight and stronger incentives to commit violations. By focusing on politically embedded SOEs, this study shows how accountability mechanisms can catalyze proactive green transformation, enhancing the strategic role of public governance in sustainability transitions.

# 1. Introduction

Despite growing global efforts—both from policymakers and the academic community—to tackle climate change, biodiversity loss, and pollution—three interconnected environmental crises identified by the United Nations Environment Programme (UNEP, 2021)—corporate environmental violations remain alarmingly widespread.[1,2] This paradox reveals a persistent governance gap, especially in contexts where internal accountability remains weak—such as in state-owned enterprises (SOEs). SOEs often dominate environmentally sensitive sectors like energy, transportation, and heavy industry. According to the 2024 OECD report, governments are increasingly expected to ensure that SOEs "lead by example" in environmental performance and sustainability disclosure.[3] Yet, many SOEs continue to lag in compliance, hindered by opaque governance and limited managerial accountability. Against this backdrop, this study examines whether institutionalized executive accountability can serve as an effective mechanism to reduce environmental violations in SOEs.

Despite its theoretical relevance, the role of internal executive accountability in curbing corporate environmental misconduct remains empirically underexplored. While prior studies have emphasized executive characteristics (Dong et al., 2024; Dong and Yu, 2024), institutional environments (Jin et al., 2024; Liu et al., 2024; Wang et al., 2024), and social oversight (Zhang et al., 2023; Yasir et al., 2024), evidence on whether formal internal accountability mechanisms can deter environmental violations is still limited. Addressing this gap, we focus on a unique institutional reform in Chinese SOEs: the "Accountability System for Irregular Operations and Investments" (ASIOI) introduced in 2016. The ASIOI policy mandates long-term accountability for managerial decisions, especially those compliance-related and major investment choices―thereby offering a unique opportunity to investigate how internal governance design fosters environmental responsibility in SOEs.

---

[1] Including the *Clean Air Act* in the United States, the *Environmental Protection Law* of China, and Japan's *Basic Environment Law*.
[2] Academic interest in environmental issues has grown considerably (e.g. Cho et al., 2010; Hofer et al., 2012; Özen and Küskü, 2009; Xu et al., 2016; Gan et al., 2019). These studies highlight that environmental violations remain pervasive, often driven by short-term growth imperatives, insufficient deterrence from penalties, and limited regulatory enforcement capacity (Marquis et al., 2011; Zhao et al., 2014; Xiong et al., 2021).
[3] Source from:
https://www.oecd.org/en/publications/ownership-and-governance-of-state-owned-enterprises-2024_395c9956-en.html.

China's rapid economic growth has led to substantial environmental costs, posing serious challenges to long-term sustainability (Zhang et al., 2023; Cheng et al., 2024). Unlike market-based approaches in Western countries, China's environmental governance emphasizes administrative control under the principle of "proactive government and effective market". Under this model, SOES, which dominate key strategic sectors―such as energy, telecommunications, transportation, and heavy industry―serve both economic and policy functions, making them central to China's green transition. However, SOEs have increasingly drawn public scrutiny for their exposure in environmental risks.[4] In addition to local environmental regulations, SOEs face additional oversight from the State-owned Assets Supervision and Administration Commission (SASAC), including the ASIOI system that aims to standardize decisions and prevent misconduct.

Since its launch in 2016, the ASIOI policy has progressed systematically under the guidance of China's State Council. By 2021, all 31 provinces and 135 prefecture-level cities had adopted ASIOI regulations. ASIOI introduced for the first time the concept of a "lifetime accountability mechanism for major decisions", clearly defining the scope of accountability across nine key domains, including control of state-owned enterprise groups, procurement and sales, risk management, and money management. This framework introduces lifetime accountability for irregularities in major investment and operational decisions, holding the involved executives both financially and legally liable (Xin et al., 2022). This unique administrative accountability regime provides a valuable empirical context for us to examine how institutional design shapes corporate environmental violations.

---

[4] Source from: https://www.nbd.com.cn/articles/2025-04-11/3828160.html;
https://www.nbd.com.cn/articles/2025-04-03/3818602.html;
https://www.nbd.com.cn/articles/2025-03-29/3811332.html.

Theoretically, this study applies the criminological cost-benefit analysis framework (Becker, 1968; Sah, 1991; Glaeser et al., 1996) to explore how ASIOI raises the cost of violations, thereby deterring environmental misconduct. Under the traditional profit-maximization logic, environmental violations are often viewed as rational choices, particularly in contexts where effective enforcement is inadequate or accountability mechanisms are unclear. ASIOI institutionalizes personal accountability and raises the costs of misconduct. By integrating ethical considerations into corporate governance, it motivates firms to pursue proactive green transformation, rather than passive compliance. Through this mechanism, ASIOI effectively aligns economic rationality with ethical responsibility, thereby establishing a balanced "financial objectives一social responsibility" framework for corporate decision-making.

Based on this institutional context, the study utilizes the staggered regional rollout of ASIOI as a quasi-natural experiment to systematically evaluate its impact on SOEs' environmental violations. The empirical results show that ASIOI significantly reduces both the probability and frequency of environmental violations. Mechanism tests indicate that the accountability system facilitates environmental compliance by enhancing internal controls, increasing green investment, and fostering green innovation. Further heterogeneity analysis reveals that the governance effect of ASIOI on corporate environmental violations is more pronounced among state-owned enterprises with weaker regulatory pressure (i.e., lower ownership by large shareholders and lower regional judicial efficiency) and stronger incentives to engage in misconduct (i.e., lower executive ownership and poorer ESG performance).

This study makes three key contributions. First, it expands the literature on the economic consequences of ASIOI, which has primarily focused on investment, auditing, and mergers and acquisitions in SOEs (e.g., Xin et al., 2022; Wu et al., 2024; Hu et al., 2025; Li and Huang, 2025; Wang et al., 2025). Little attention has been paid to its environmental governance effects. By incorporating ASIOI into the cost-benefit analysis framework, this paper demonstrates how administrative institutional tools can effectively constrain corporate environmental misconduct, thereby extending the research boundaries on administrative institutions.

Second, it enriches our understanding of what drives corporate environmental violations. While prior studies have primarily focused on the role of environmental policies and regulatory enforcement (e.g., Jin et al., 2024; Liu et al., 2024; Wang et al., 2024), limited attention has been paid to the role of executive accountability systems. Thus, this study contributes to a more comprehensive understanding of the determinants of environmental misconduct.

Finally, this study identifies the moderating roles of supervisory intensity (proxied by large shareholder ownership and judicial efficiency) and misconduct motivation (proxied by executive ownership and ESG scores) in shaping the regulatory effects of the system. These findings deepen our understanding of the sources of heterogeneity in policy implementation outcomes and provide a theoretical foundation for more targeted policy design.

The remainder of this study is structured as follows: Section 2 introduces the institutional background, literature review, and research hypotheses. Section 3 details the research design, including data sources, variable measurement, and model specification. Section 4 presents the empirical results, followed by robustness checks, mechanism and heterogeneity analyses. Section 5 concludes the study and offers policy recommendations.

## 2. Institutional Background, Literature Review and Hypothesis Development

### 2.1 Accountability System for Illegal Operation and Investment of State-owned Enterprises in China (ASIOI)

Over the years, the CPC Central Committee and the State Council have attached great importance to the accountability for the loss of state-owned assets. The 2006 Government Work Report at the Fourth Session of the 10th National People's Congress emphasized the need to improve the supervision system for state-owned assets and establish an accountability mechanism for significant losses. In 2008, to strengthen the protection of state-owned assets and regulate accountability procedures, the SASAC issued the "Interim Measures for the Accountability of Asset Losses in Central SOEs". This marked China's initial establishment of an accountability system for business operations and investments in SOEs. However, it applied only to central SOEs and lacked broader applicability and refinement, limiting nationwide implementation.[5]

Since the 18th National Congress of the CPC, President Xi Jinping has repeatedly underscored the importance of a sound accountability framework. In 2013, the Third Plenary Session of the 18th CPC Central Committee proposed a long-term incentive and restraint mechanism and reinforced the need for accountability in SOE operations and investments. In the same year, to enhance accountability among corporate leaders and protect state-owned assets, the SASAC of Shijiazhuang City issued the "Interim Measures for the Accountability of Business Operations and Investments of SOE Executives in Shijiazhuang".[6]

---

[5] Source from https://www.np.gov.cn/cms/html/npszf/2009-11-22/262226090.html.

Subsequently, in 2015, the "Guiding Opinions on Deepening the Reform of State-Owned Enterprises", issued by the CPC Central Committee and the State Council, explicitly identified strict accountability as a key reform priority.[7] In October of the same year, the State Council released the "Opinions on Reforming and Improving the State-Owned Assets Management System", which called for establishing a sound accountability framework for irregular operations and investments in SOEs, as well as a retrospective accountability mechanism for major decision-making errors and misconduct. In line with this direction, Zhejiang Provincial Government issued the "Interim Measures for the Accountability of Asset Losses from Business Operations and Investments in Provincial SOEs" to safeguard state-owned assets and mitigate investment risks.[8]

However, the above frameworks developed before 2015 have not yet institutionalize the concept of "lifetime accountability for major decisions". To fill this gap, China's State Council General Office issued the "Opinions on Establishing an Accountability System for Irregular Operations and Investments in SOEs" in 2016. This document formally introduced the concept of lifetime accountability for major decisions and mandated that, by 2020, a comprehensive ASIOI framework be established across all levels of institutions fulfilling state capital ownership functions.[9] The system clearly defines the scope of accountability across nine key domains, including control of state-owned enterprise groups, procurement and sales, risk management, and money management. It also defined asset loss categories—general and significant—based on local conditions and assigned responsibilities into three types: direct, supervisory, and leadership.

---

[6] Source from http://www.sasac.gov.cn/n2588025/n2588129/c2711001/content.html.
[7] Source from https://www.gov.cn/gongbao/content/2015/content_2937313.html.
[8] Source from https://www.zj.gov.cn/art/2023/1/4/art_1229017138_2453907.html.
[9] Source from https://www.gov.cn/zhengce/content/2016-08/23/content_5101590.html.

To ensure implementation, SASAC established two dedicated supervisory departments—Supervision Bureaus II and III—in November 2016 to handle matters related to ASIOI. By focusing on ex post accountability and lifetime liability for major decisions, the ASIOI framework represents an institutional innovation in state capital oversight, enhancing long-term regulatory control over SOE investments.[10]

As shown in Figure 1, since 2014, accountability frameworks for business operations and investments have gradually been established at the provincial and municipal levels. In 2014 and 2015, only a few cities had introduced such systems. However, after the concept of lifetime accountability for major decisions was introduced in 2016, the number of these frameworks increased significantly. By 2017, 17 provinces had adopted such systems, along with 23 municipalities. From 2018 onward, municipal-level adoption surged, reaching 135 cities by 2021. By the same year, all 31 inland provinces had implemented provincial-level accountability frameworks. This trend reflects the systematic and nationwide diffusion of ASIOI, signifying a major acceleration in the institutionalization and standardization of SOE governance.

Between 2016 and 2021, the decentralized establishment of ASIOI mechanisms across regions created a favorable quasi-natural experimental setting. The development of ASIOI aims to draw upon past lessons to build a clearer and more effective accountability system for SOE operations and investments, thereby supporting the long-term, sustainable development of state-owned enterprises.

---

[10] Following the national-level guidance, local governments successively introduced supporting regulations. In October 2016, the Yancheng SASAC issued the "Interim Measures for the Accountability of Operational Errors and Asset Losses in Municipal SOEs", applicable upon release. In December, the Anhui Provincial SASAC released the "Interim Measures for the Accountability of Irregular Operations and Investments in Provincial SOEs", effective from 2017. In July 2018, SASAC issued the "Trial Implementation Measures for the Accountability of Irregular Operations and Investments in Central SOEs", which came into effect in August 2018.

*/\* Insert Figure 1 here \*/*

**2.2 Literature Review**

*2.1.1 Research on the Economic Consequences of ASIOI*

Since its formal implementation in 2016, ASIOI has attracted significant attention from practitioners for its potential to reshape the governance of SOEs. While scholarly research on ASIOI is still emerging, a growing body of literature has begun to examine its implications, particularly its impact on SOE investment behavior, auditing, and mergers and acquisitions.

Specifically, prior research generally finds that ASIOI helps curb short-termism in SOEs and promotes their development toward more standardized and long-term orientations. For instance, Xin et al. (2022) show that ASIOI improves internal control quality and mitigates managerial short-sightedness, thereby fostering more disciplined operations in SOEs. Similarly, Wang et al. (2025) further demonstrate that the system enhances investment efficiency by reducing agency costs and improving information flow.

In addition, Hu et al. (2025), from the perspective of financial asset allocation, find that ASIOI curbs excessive and speculative financial investments, indicating its governance effect in guiding rational resource allocation. Wu et al. (2024), adopting an auditing perspective, argue that ASIOI reduces earnings management and misconduct, thereby improving internal control and audit quality in SOEs. Li and Huang (2025) further find that ASIOI suppresses irrational mergers and enhances post-merger integration effectiveness.

*2.2.2 Research on the Determinants of Environmental Violations*

A growing body of literature has explored the determinants of corporate environmental violations, which can be categorized into managerial characteristics, institutional factors, and social and technological forces.

Prior research underscores the pivotal role of executive background and incentives in shaping environmental outcomes. Specifically, the environmental management and overseas experience help improve a firm's environmental governance capacity, thereby reducing the risk of environmental violations (Dong et al., 2024; Dong and Yu, 2024).

Externally imposed regulatory policies form another critical line of defense against environmental violations. For instance, Jin et al. (2024) show that the environmental tax reform—by increasing local tax revenues and incentivizing firms to invest in environmental protection—significantly reduces environmental violations. Wang et al. (2024) find that China's low-carbon city pilot policy, by strengthening governmental focus on green development and increasing environmental subsidies and investments, effectively reduces violations by heavily polluting firms. Liu et al. (2024) further highlight the complementary role of formal and informal regulation in generating effective institutional pressure, thereby deterring environmental misconduct.

The enhancement of social supervision and technological progress have also promoted the improvement of environmental compliance. Zhang et al. (2023) find that public involvement in environmental governance helps promote stricter local environmental regulations, which in turn reduce corporate violations. Yasir et al. (2024) find that social trust can also reduce further environmental violations of enterprises. Fahad et al. (2024) point out that digital finance, as a technological tool, reduces environmental violations by improving internal controls and enhancing environmental investment. This effect is especially pronounced in SOEs, heavily polluting industries, and regions with stringent environmental regulations.

Overall, existing studies suggest that ASIOI can guide firms toward long-term development by improving corporate governance and enhancing internal controls. However, current literature mainly focuses on ASIOI's impact on operational outcomes such as investment efficiency, audit quality, and mergers and acquisitions, with limited attention to its role in environmental governance. Meanwhile, research on environmental violations has largely explored managerial traits, institutional environments, and social oversight, identifying various important deterrents, yet largely overlooking the role of ASIOI. Building on this foundation, this study seeks to examine whether, and through what mechanisms, ASIOI influences environmental violations in SOEs.

**2.3 Hypothesis Development**

Classical theories of economic crime emphasize the rational calculation underlying rule violations. Becker (1968) constructed a foundational cost–benefit analysis model, which posits that an individual's decision to engage in illegal behavior depends on the trade-off between the expected benefits of the behavior and the potential costs of violation. This theory emphasizes that when the expected gains from illegality outweigh the associated costs, individuals are more likely to break the law.

Subsequent studies (e.g., Sah, 1991; Glaeser et al., 1996) further expanded this model, arguing that the institutional environment, regulatory intensity, and individuals' risk aversion all play crucial roles in shaping decision-making. When this analytical framework is applied to corporate environmental compliance decisions, it suggests that a firm's inclination to commit environmental violations is also constrained by the strategic balance between potential benefits (e.g., cost savings from non-compliance, short-term profit maximization) and violation costs (e.g., probability and severity of penalties, scope and depth of accountability).[11]

When applied to corporate environmental behavior, this framework suggests that firms will tend to comply with environmental regulations only when the perceived costs of violation—such as penalties, reputational damage, or career consequences—exceed the short-term economic benefits of noncompliance (Karpoff et al., 2005). However, in many developing institutional countries[12], a persistent dilemma arises: firms often face high compliance costs but low expected penalties, leading to widespread environmental violations and a disregard for ecological accountability (Zhao et al., 2014; Xiong et al., 2021).

To break this "low-cost violation trap", China's SASAC introduced the "Accountability System for Irregular Operations and Investments" (ASIOI) in 2016. The system mandates lifetime and retrospective accountability for SOE executives who make significant misjudgments or engage in misconduct. Unlike conventional punitive mechanisms targeting the legal entity, ASIOI directly assigns personal accountability to decision-makers—even after they leave office—thereby shifting the cost–benefit calculus of corporate behavior. [13]

From the criminological cost–benefit analysis perspective, ASIOI significantly increases the cost of corporate violations through three dimensions. First, accountability is directed at individual executives, extending beyond the corporate entity of the enterprise to the individual decision-makers, thus breaking the institutional inertia of "corporate violations, personal impunity" (Xin et al., 2022).

---

[11] Recent research by Dong et al. (2018) in the *Journal of Business Ethics* highlights how institutional arrangements—such as social trust—shape the perceived costs of misconduct and influence firm behavior.
[12] In the absence of effective regulation and accountability mechanisms.
[13] The scope of disciplinary measures includes public criticism, pay cuts, suspension, dismissal, and even industry bans. Importantly, accountability is not limited to an official's tenure, as the principle of "lifetime responsibility" applies—greatly strengthening behavioral constraints on SOE executives (Xin et al., 2022).

Second, the non-monetary costs resulting from accountability—such as restricted career progression, reputational damage, and loss of future employment opportunities—are far greater than traditional economic penalties, heightening executives' psychological expectations of adverse consequences (Xin et al., 2022). Third, in the process of enforcement, mechanisms such as "retrospective investigation" and "public disclosure of accountability" raise the likelihood of violations being detected and punished, thereby systematically increasing the cost of misconduct (Sah, 1991; Glaeser et al., 1996). Under such a system, SOE executives are more inclined to adopt more responsible and sustainable behavioral patterns when faced with environmental decisions that may have significant consequences.

Accordingly, this study proposes the following research hypothesis:

Hypothesis 1: The implementation of ASIOI significantly reduces environmental violations by SOEs.

## 3. Research design

### 3.1 Sample selection

This study selects A-share listed companies on the Shanghai Stock Exchange and Shenzhen Stock Exchange in China, focusing on heavily polluting industries during the period from 2013 to 2021 as the research sample. The timing of ASIOI implementation by central and local governments is manually collected. Data on environmental violations are obtained by matching environmental violation events disclosed by environmental protection bureaus with listed company information. Other firm-level financial data are sourced from the CSMAR and CNRDS databases, both of which are leading financial data providers in China.

Sample selection follows these criteria: (1) excluding observations from the financial industry; (2) excluding observations with a leverage ratio greater than 1; (3)

excluding companies under "special treatment" (ST and *ST companies); (4) excluding observations with missing values for key variables. To mitigate the influence of outliers on empirical results, all continuous variables used in regressions are winsorized at the 1st and 99th percentiles. The final sample consists of 6,718 observations.

**3.2 Key Variable Definitions**

Following existing studies (Dong and Yu, 2024; Dong et al., 2024), this study measures corporate environmental violations (ENV) using both the probability and frequency of violations. To measure the probability of environmental violations, we construct the variable ***DUM_ENV***, which equals 1 if a state-owned enterprise committed an environmental violation in a given year, and 0 otherwise. To measure the frequency of environmental violations, we construct ***FRE_ENV***, defined as the natural logarithm of the number of environmental violation events in a given year plus one.

In line with Xin et al. (2022), this study manually collects the policy issuance dates of operational investment accountability systems at the central and regional levels, constructing policy dummy variables based on implementation timing. Policy information is obtained from official government documents, announcements, notifications, and related news reports. Consistent with existing literature (Xin et al., 2022), the 2016 release of the "Opinions on Establishing a System for Accountability of Irregular Operations and Investments in State-Owned Enterprises" is considered the starting point for the systematic establishment of the ASIOI framework. Therefore, we exclude samples from Shijiazhuang's municipal SOEs and Zhejiang's provincial SOEs to avoid biasing the results. The core explanatory variable ***DID*** is a dummy variable. For local SOEs, it is equal to 1 if ASIOI is implemented in the location of the SOEs in a given year, and 0 otherwise. For central SOEs, as they have uniformly implemented ASIOI since 2018, the value is 1 if the year is 2018 or later, and 0 otherwise. Definitions of other variables are provided in Appendix 1.

**3.3 Model Specification**

This study constructs Model (1) to test the impact of ASIOI implementation on corporate environmental violations.

$$\text{ENV}_{it} = \alpha_0 + \alpha_1 \text{DID} + \sum_{k}^{n} a_k \text{Controls}_{it} + \mu_i + \gamma_{jt} + \epsilon_{it} \qquad (1)$$

In this model, the subscript i denotes the firm, t denotes the year, and j denotes the industry. The dependent variable is corporate environmental violations (*ENV*), measured using two indicators: the probability of environmental violations (*DUM_ENV*) and the frequency of environmental violations (*FRE_ENV*). This study employs a multi-period difference-in-differences (DID) model, where the key explanatory variable is *DID*. If a state-owned listed company is located in a region that implemented the ASIOI policy in a given year, then *DID* = 1; otherwise, *DID* = 0. This variable captures the difference in environmental violations between firms affected by the policy and those not affected, similar to the interaction term in a traditional DID model. The key coefficient in this model captures the effect of ASIOI on corporate environmental violations.

Additionally, based on existing literature (Dong and Yu, 2024; Wang et al., 2024; Jin et al., 2024), this study incorporates a series of firm-level control variables into the empirical regressions, including: firm age (*Age*), firm size measured as the natural logarithm of total assets (*Size*), leverage ratio measured by the ratio of total liabilities to total assets (*Lev*), return on assets measured as net profit over total assets (*ROA*), net cash flow from operating activities (*Cashflow*), CEO duality (*DUAL*), board size (*Boardsize*), ownership concentration measured by the shareholding ratio of the largest shareholder (*Top1*), and proportion of independent directors (*IndRatio*).

Various fixed effects are included in the regressions to account for omitted variables that may correlate with *DID* or affect environmental violations. On one hand, firm fixed effects ($\mu_i$) are included to control for time-invariant firm characteristics. On the other hand, year fixed effects ($\gamma_{jt}$) are introduced to account for time-specific factors affecting all firms, such as annual policy changes and market conditions. In addition, the standard errors of all regression coefficients are clustered at the industry level to improve the robustness of the estimates.

## 4. Empirical Analysis

### 4.1 Descriptive Statistics

Table 1 reports the descriptive statistics for the main variables. The mean of corporate environmental violation probability (*DUM_ENV*) is 0.048 with a standard deviation of 0.215. The mean of environmental violation frequency (*FRE_ENV*) is 0.054 with a standard deviation of 0.276, indicating significant variation in environmental violations among firms. The policy variable (*DID*) has a mean of 0.172, suggesting that 17.2% of the observations were affected by the policy. The mean of the largest shareholder's ownership (*Top1*) is 34.932%, with a standard deviation of 14.784%, indicating a relatively high degree of ownership concentration. Overall, the descriptive statistics of the variables are consistent with existing literature (Dong and Yu, 2024; Wang et al., 2024; Jin et al., 2024), confirming the appropriateness of the sample.

*/* Insert Table 1 here */*

## 4.2 Baseline Regression Results

To examine the impact of ASIOI on corporate environmental violations, regressions are conducted based on Model (1). The results are presented in Table 2. In Column (1), the regression coefficient of **DID** is -0.0295 and statistically significant at the 5% level. In Column (2), after including a set of control variables, the **DID** coefficient remains statistically significant at -0.0281. This implies that, relative to the control group, the implementation of the ASIOI policy leads to a 2.81 percentage point reduction in the probability of environmental violations among treated firms. In Column (3), the **DID** coefficient is -0.0379 and significant at the 5% level; after adding control variables in Column (4), the coefficient is -0.0365, also significant at the 5% level. Given that **FRE_ENV** is log-transformed, and we use log(1 + number of violations) to retain observations with zero violations, the coefficient can be interpreted approximately as a 3.65% average decrease in the frequency of violations among treated firms, relative to the control group.

The baseline regression results suggest that ASIOI policy significantly reduces both the probability and frequency of corporate environmental misconduct, providing preliminary support for Hypothesis 1 (H1).

/* Insert Table 2 here */

## 4.3 Robustness Tests

*4.3.1 Parallel Trend Test*

Since the main empirical strategy is based on the DID model, a parallel trend test is required to verify that the treatment and control groups followed similar trends before the implementation of ASIOI (Wang et al., 2024; Jin et al., 2024). Specifically, the following Model (2) is designed to test the parallel trend assumption.

$$ENV_{it} = \alpha + \beta_1 Before^{-3} + \beta_2 Before^{-2} + \beta_3 Current + \beta_4 After^1 + \beta_5 After^2 + \beta_6 After^3 + \sum_{k}^{n} a_k Controls_{it} + \mu_i + \gamma_{jt} + \epsilon_{it} \qquad (2)$$

In this model, ***Before$^{-3}$*** is a dummy variable equal to 1 if the year is three years before (or earlier than) the policy implementation; ***Before$^{-2}$*** equals 1 if the year is two years before implementation. ***Current*** is a binary variable equal to 1 if it is the year of policy implementation, indicating the observation year of the policy. ***After$^1$***, ***After$^2$***, and ***After$^3$*** are dummy variables equal to 1 if the year is one, two, and three or more years after the policy implementation, respectively. ***Before$^{-1}$*** is used as the reference group, representing the year before policy implementation.

As shown in Table 3, the dummy variables for the years before policy implementation (***Before$^{-3}$***, ***Before$^{-2}$***) have no significant effect on environmental violations (***DUM_ENV***, ***FRE_ENV***), indicating that the treatment and control groups followed parallel trends in environmental misconduct before the policy was enacted. The effect of ***Current*** on environmental violations is statistically insignificant at the 10% level, suggesting that the policy's immediate effect was relatively weak. After policy implementation (***After$^1$***, ***After$^2$***, ***After$^3$***), the dummy variables have significantly negative coefficients for ***DUM_ENV*** at the 5% significance level. Although ***After$^1$*** has an insignificant effect on ***FRE_ENV***, the coefficients for ***After$^2$*** and ***After$^3$*** are significantly negative at the 5% or 10% levels, respectively.

These results indicate that ASIOI significantly curbs corporate environmental violations after implementation, and the effect strengthens over time. This supports the parallel trend assumption and demonstrates that the accountability system's impact becomes more pronounced as the policy takes root, further supporting H1.

*/* Insert Table 3 here */*

*4.3.2 Placebo Test*

The placebo test is an important method for assessing the robustness of causal inference. By randomly generating policy implementation years and repeating the experiment multiple times, one can detect whether the model suffers from systematic bias. If the randomized policy variable has no significant effect, it confirms the robustness of the original model's causal inference. Following existing literature (Wang et al., 2024; Jin et al., 2024), this study randomly assigns ASIOI implementation years using a computer algorithm to generate a new policy variable and examines whether it still suppresses environmental violations.

As shown in Figures 2, the randomization process is repeated 1,000 times. The coefficients of the randomly assigned policy variables are centered around zero and show no significant deviation, indicating that ASIOI's impact on environmental violations disappears under random assignment. Moreover, the coefficients are much larger than those in the baseline regression. This confirms that the negative effect of ASIOI on corporate environmental violations found in the baseline regression reflects a true causal relationship, further validating the robustness of the findings.

*/* Insert Figure 2 here */*

*4.3.3 Change the Regression Model*

Given that corporate environmental violations (**DUM_ENV**) are represented as a binary variable (0-1), this study replaces the linear regression model with a logit model in the robustness check to enhance the credibility of the results. Table 4 reports the estimation results of the logit regressions. In column (1), without controlling for firm-level characteristics, the coefficient of **DID** is -0.8673 and statistically significant at the 1% level. In column (2), after controlling for a series of firm-specific variables, the coefficient of ***DID*** is -0.9264, also significant at the 1% level. These results suggest that even when using a nonlinear probability model, the negative effect of ASIOI on corporate environmental violations remains robust, further confirming the validity of the baseline regression findings.

/* Insert Table 4 here */

*4.3.4 CSDID*

To address the potential heterogeneity in multi-period DID models, this study adjusts the model specification and re-estimates the effects. Prior research (Goodman-Bacon, 2021) has highlighted that staggered treatment timing may result in early-treated units being incorrectly used as control groups for later-treated units, which can lead to biased estimates. Following the approach of Callaway and Sant'Anna (2021), we adopt the CSDID method to eliminate such invalid comparisons and improve the accuracy of the estimates. As shown in Table 5, the results from the CSDID model are consistent with those from the baseline regression.

/* Insert Table 5 here */

*4.3.5 Changing the Clustering Level*

To enhance the robustness of the results, the standard errors of all regression coefficients are clustered at the firm level. As shown in Table 6, in Column (1), the coefficient of ***DID*** is -0.0295 and significant at the 5% level. In Column (2), after including control variables, the coefficient is -0.0281, also significant at the 5% level. In Column (3), the ***DID*** coefficient is -0.0379 and significant at the 1% level; after adding control variables in Column (4), the coefficient is -0.0365, significant at the 5% level. These results indicate that the suppressive effect of ASIOI (***DID***) on environmental violations (***DUM_ENV***, ***FRE_ENV***) remains significant even when the clustering level is changed.

*/* Insert Table 6 here */*

*4.3.6 Changing the Fixed Effects*

Furthermore, this study controls for firm fixed effects and industry-year fixed effects to exclude the influence of unobserved macroeconomic conditions, local policies, and regional factors across industries and years. This allows for more accurate identification of the policy's effect. In the regression results reported in Table 7, Column (1) shows that the **DID** coefficient is -0.0320, significant at the 5% level. After adding control variables in Column (2), the coefficient becomes -0.0325, also significant at the 5% level. In Column (3), the **DID** coefficient is -0.0389, significant at the 5% level; and in Column (4), after including control variables, it is -0.0396, also significant at the 5% level. These findings demonstrate that the suppressive effect of ASIOI (**DID**) on corporate environmental violations (**DUM_ENV**, **FRE_ENV**) remains significant even when alternative fixed effects are used.

*/* Insert Table 7 here */*

**4.4 Mechanism test**

The results in the previous section indicate that the implementation of ASIOI has had a positive impact on corporate environmental violations. However, further tests are required to determine how ASIOI precisely reduces such violations. From the perspective of criminology's cost-benefit theory, ASIOI significantly increases the potential penalty costs of environmental violations for enterprises, prompting SOEs to gradually shift toward a decision-making path that emphasizes long-term returns and risk prevention, rather than merely pursuing short-term gains. SOE managers have gradually strengthened their sense of responsibility and awareness of risk prevention, no longer being solely profit-driven in the short term. Instead, they have begun to consider environmental governance from a long-term perspective, incorporating compliance and sustainable development into corporate strategic considerations.

Different from the traditional "deterrence–evasion" logic (Xin, 2022), ASIOI not only exerts external pressure but also guides SOEs to proactively reshape their development paths through institutional arrangements, demonstrating stronger foresight and responsibility orientation. Specifically, enterprises adjust their environmental governance strategies and achieve a transition from "passive compliance" to "proactive green transformation" through three primary mechanisms: improving internal governance structures, increasing green investment, and promoting green innovation.

*4.4.1 Internal control*

Effective internal control can serve as a safeguard to ensure transparent reporting and ethical conduct (Fernandhytia and Muslichah, 2020), thereby reducing the likelihood of environmental violations. Referring to existing research (Fahad et al., 2024), this study employs the DIB Internal Control Index of China to quantify internal control (**IC_Index**).

As shown in Table 8, in column (1), the coefficient of **DID** on internal control (**IC_Index**) is 0.1500 and is significant at the 10% level. In column (2), after adding control variables, the coefficient of **DID** on internal control (**IC_Index**) is 0.1153 and remains significant at the 10% level. These results suggest a significant positive relationship between the implementation of ASIOI and the improvement in internal control quality.

*/\* Insert Table 8 here \*/*

*4.4.2 Green investment*

Enterprises can reduce environmental violations by actively increasing the scale of green investment (Chen and Ma, 2021). In this study, green investment (*GInvest*) is measured as the logarithm of one plus the amount of green investment in the given year (Jin et al., 2024).

As shown in Table 9, in column (1), the coefficient of **DID** on green investment (*GInvest*) is 0.9320 and is significant at the 10% level. In column (2), after adding control variables, the coefficient increases to 1.2205, with improved significance. These results indicate a significant positive relationship between the implementation of ASIOI and the enhancement of corporate green investment.

*/* Insert Table 9 here */*

*4.4.3 Green innovation*

Green innovation reflects a firm's core technological capability and knowledge output in environmental sustainability. Active participation in green innovation can reduce corporate environmental violations (Jin et al., 2024). This study measures green innovation using the logarithm of one plus the number of green patent applications filed annually.

As shown in Table 10, in column (1), the coefficient of **DID** on green innovation (*Innovation*) is 0.1105 and is significant at the 5% level. In column (2), after adding control variables, the coefficient is 0.1011 and remains significant at the 5% level. Considering that green innovation takes time to accumulate, we further regress green innovation with a one-period lag. As shown in columns (3) and (4) of Table 9, the coefficient of **DID** on lagged green innovation (*F.Innovation*) remains significantly positive. These results suggest a significant positive relationship between the implementation of ASIOI and the improvement in corporate green innovation.

*/* Insert Table 10 here */*

## 4.5 Heterogeneity test

The effectiveness of ASIOI in curbing environmental violations of state-owned enterprises (SOEs) may depend on the intensity of supervision and the motives behind these violations. From the perspective of supervision, SOEs with weak governance structures or those located in regions with weak enforcement are less likely to be effectively monitored through conventional channels. Under such circumstances, ASIOI, as a targeted accountability mechanism, can play a stronger role. In terms of the motives for violation, SOEs with executives lacking long-term incentives or poor environmental governance are more likely to pursue short-term profits at the expense of environmental compliance. In these cases, the accountability pressure brought by ASIOI is also likely to generate a stronger governance effect.

Based on this, the present study conducts a heterogeneity analysis across four dimensions: (1) the shareholding ratio of the largest shareholder; (2) regional judicial efficiency; (3) the shareholding ratio of executives; and (4) ESG score. The results confirm our hypothesis that ASIOI has a more pronounced inhibitory effect on environmental violations of SOEs in companies with weaker supervision and stronger motives for violation.

### 4.5.1 Oversight Strength

#### 4.5.1.1 Shareholding ratio of the largest shareholder

Ultimate controlling shareholders are typically the final decision-makers in the companies they own (Jiang & Kim, 2020). Violations of environmental regulations can provoke strong public opposition and damage the reputation of these controlling shareholders. In addition, the controlling shareholder's wealth is closely tied to the firm, which gives them stronger incentives to monitor corporate behavior. We predict that in firms with a lower shareholding ratio of the largest shareholder—indicating weaker oversight—the restraining effect of ASIOI on corporate environmental violations will be stronger.

Therefore, we further divide the sample according to the shareholding ratio of the largest shareholder to examine its moderating effect on the relationship between ASIOI and corporate environmental violations.

Specifically, firms with a shareholding ratio of the largest shareholder above the annual median are classified as *Top1_High*, and those below the median as *Top1_Low*. As shown in Table 11, for the Top1_Low group, the **DID** coefficients on both the probability of environmental violations (**DUM_ENV**) and the frequency of violations (**FRE_ENV**) are significantly negative at the 5% level. This suggests that ASIOI significantly reduces environmental violations in SOEs where the shareholding ratio of the largest shareholder is relatively low. In contrast, in the Top1_High group, the **DID** coefficients for **DUM_ENV** and **FRE_ENV** are not statistically significant at the 10% level, indicating that in SOEs with a high shareholding ratio of the largest shareholder, the deterrent effect of the accountability system for environmental violations is relatively weak.

*/* Insert Table 11 here */*

*4.5.1.2 Regional Judicial Efficiency*

Previous research suggests that improvements in the quality of environmental justice significantly deter corporate environmental violations (Yu et al., 2024). In regions with low judicial efficiency, where the enforcement of environmental regulations is weaker and the cost of violations is lower, ASIOI serves as a critical institutional supplement. It imposes targeted accountability on SOE executives, thereby enhancing compliance where external legal deterrence is lacking.

This study employs the Market Intermediary and Legal Environment Index from Wang et al. (2017) Marketization Index to measure regional judicial efficiency. We further group the sample based on judicial efficiency to evaluate its moderating effect on the ASIOI–environmental violation relationship.

Specifically, regions with judicial efficiency above the annual median are classified as ***Judicial efficiency_High***, and those below as ***Judicial efficiency_Low***. As shown in Table 12, for the Judicial efficiency_Low group, the coefficient of ***DID*** on ***DUM_ENV*** is significantly negative at the 5% level, and the coefficient on ***FRE_ENV*** is significantly negative at the 1% level. This indicates that ASIOI significantly inhibits environmental violations in SOEs located in regions with low judicial efficiency. In contrast, for the Judicial efficiency_High group, the coefficients of ***DID*** on both ***DUM_ENV*** and ***FRE_ENV*** are not statistically significant at the 10% level, suggesting that the effectiveness of ASIOI in curbing environmental violations is weaker in regions with high judicial efficiency.

*/* Insert Table 12 here */*

*4.5.2 Violation Motivation*

*4.5.2.1 Executive Shareholding*

The proportion of ownership held by management is closely associated with outcomes such as corporate green innovation and environmental governance behaviors (Han & Luo, 2024). Boards of directors are established to represent investors and oversee corporate management. Executives with higher shareholding ratios tend to have interests more closely aligned with the long-term goals of the firm. In contrast, those with lower shareholding ratios may lack long-term incentives and are thus more inclined to pursue short-term gains in environmental management, exhibiting stronger motivations for environmental violations.

Prior studies (Dahya et al., 2002; Weisbach, 1988) have shown that boards can serve as effective monitors. In this study, the chairman's shareholding ratio is used as the proxy variable for executive shareholding. Accordingly, we further conduct a grouped regression based on executive shareholding ratios to examine their moderating effect on the relationship between ASIOI and corporate environmental violations.

Specifically, firms with executive shareholding ratios above the annual industry median are classified as ***Executive Shareholding_High***, while those below the median are classified as ***Executive Shareholding_Low***. As shown in Table 13, for the Executive Shareholding_Low group, the ***DID*** coefficients for both the probability of environmental violations (***DUM_ENV***) and the frequency of violations (***FRE_ENV***) are significantly negative at the 5% level. This indicates that ASIOI significantly restrains environmental violations in SOEs with lower executive shareholding. In contrast, for the Executive Shareholding_High group, the ***DID*** coefficients for ***DUM_ENV*** and ***FRE_ENV*** are not statistically significant at the 10% level, suggesting that in SOEs with higher executive shareholding, the deterrent effect of ASIOI on environmental violations is relatively weak.

*/\* Insert Table 13 here \*/*

*4.5.2.2 ESG score*

Good ESG performance helps enterprises gain recognition and support from stakeholders (Martin and Moser, 2016), secure competitive resources (Flammer, 2015), and accumulate reputational capital, which is vital for sustainable development (Dhaliwal et al., 2011; Goss and Roberts, 2011). Existing literature shows that ESG performance attracts stakeholder attention, reduces information asymmetry (Eliwa et al., 2019), and improves both internal and external monitoring mechanisms (Liu and Lin, 2022), thereby weakening the motivation and justification for corporate environmental violations. However, those enterprises with lower ESG scores and weaker green governance concepts have higher motivation for environmental violations, and ASIOI as an external mechanism may play a more effective role.

This study measures ESG performance using ESG scores from the CNRDS database (Liu et al., 2024). A grouped analysis is conducted to assess the moderating effect of ESG performance on the relationship between ASIOI and environmental violations.

Specifically, firms with ESG scores above the annual median are defined as **Score_High**, and those below as **Score_Low**. As shown in Table 14, for the Score_Low group, the coefficients of **DID** on **DUM_ENV** and **FRE_ENV** are both significantly negative at the 5% level, indicating that ASIOI significantly restrains environmental violations in SOEs with low ESG scores. In contrast, in the Score_High group, the coefficients of **DID** on **DUM_ENV** and **FRE_ENV** are not statistically significant at the 10% level, suggesting a weaker disciplinary effect of ASIOI in SOEs with high ESG performance.

/* Insert Table 14 here */

## 5. Conclusion

This study investigates the governance effects of China's "Accountability System for Irregular Operations and Investments" (ASIOI) on environmental violations by state-owned enterprises (SOEs). The findings show that the ASIOI significantly reduces both the probability and frequency of environmental violations among SOEs. This result confirms the effectiveness of institutionalized ethical norms in constraining corporate behavior and provides empirical support for the transition of SOEs from "passive compliance" to "proactive green transformation." The findings remain robust across various tests. The analysis further reveals that the ASIOI influences corporate environmental behavior through three primary channels: strengthening internal controls, increasing green investment, and promoting green innovation. The heterogeneity test shows that the governance effect is more obvious in SOEs with lower supervision and stronger incentive to violate regulations.

From a theoretical perspective, this study introduces the criminological cost-benefit analysis framework into the study of corporate environmental violations. It demonstrates how institutional accountability mechanisms influence organizational behavior by raising the cost of violations and reinforcing personal responsibility. The research contributes to the literature on corporate ethics by addressing a key question: how institutional design can improve corporate social responsibility. It highlights the reinforcing relationship between formal institutional constraints and moral responsibility.

This study also offers important policy implications. First, for developing countries, especially those with a high proportion of public ownership, relying solely on market discipline and conventional regulatory enforcement is insufficient for effective environmental governance. Enhancing accountability mechanisms and improving decision-making transparency are essential to strengthening corporate ethical constraints. Second, the effectiveness of the ASIOI illustrates that embedding clear ethical responsibility pathways in corporate governance can guide firms beyond short-term profit motives toward greater attention to their environmental and social impacts. Third, institutional design should account for corporate heterogeneity by implementing dynamic regulation and layered accountability frameworks to ensure fairness, adaptability, and a balance between regulatory efficiency and enterprise development.

In sum, the ASIOI represents a key milestone in China's SOE governance reform and offers the international community a novel model of corporate sustainability grounded in an "accountability–compliance–ethics" pathway. This model underscores the institutional and moral value of formalizing ethical responsibility and highlights its critical role in encouraging firms to take proactive responsibility in environmental governance and achieve green transformation.

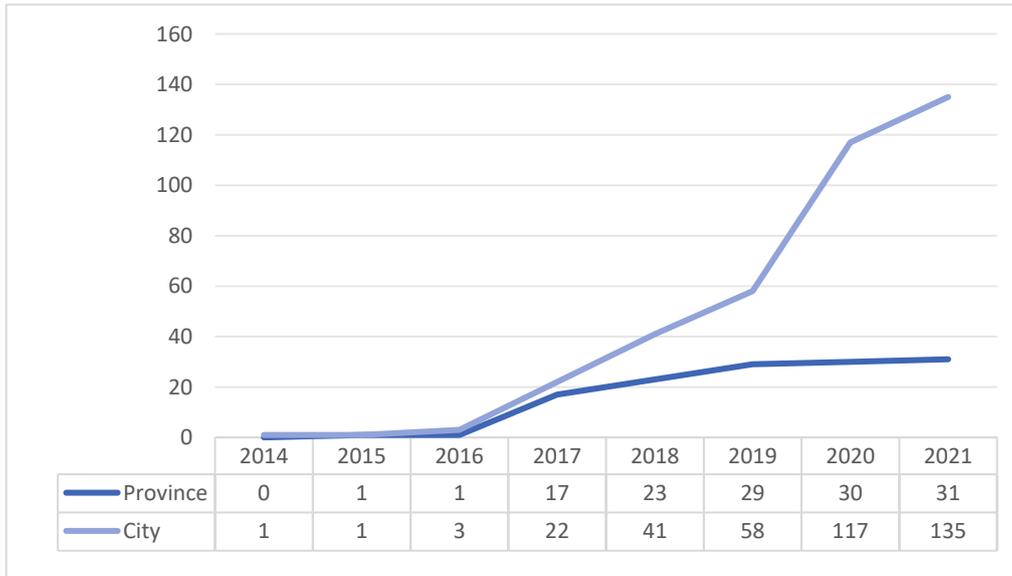

**Figure 1. Cumulative Implementation of ASIOI Across Provinces and Prefecture-Level Cities[14]**

---

[14] Specifically, the policy implementation for municipally owned state-owned enterprises (SOEs) in Shijiazhuang City is recorded as 2014, based on local government reports. For provincially owned SOEs in Zhejiang Province, the official announcement was made in 2015. The implementation time for centrally administered SOEs is uniformly set as 2018.

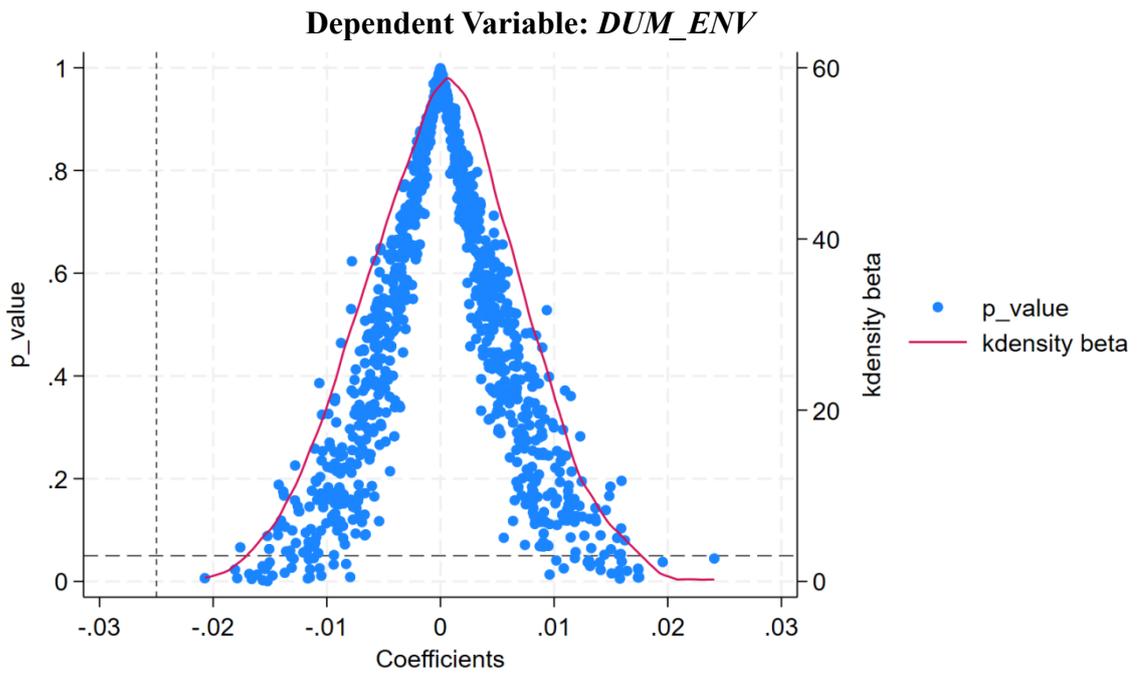

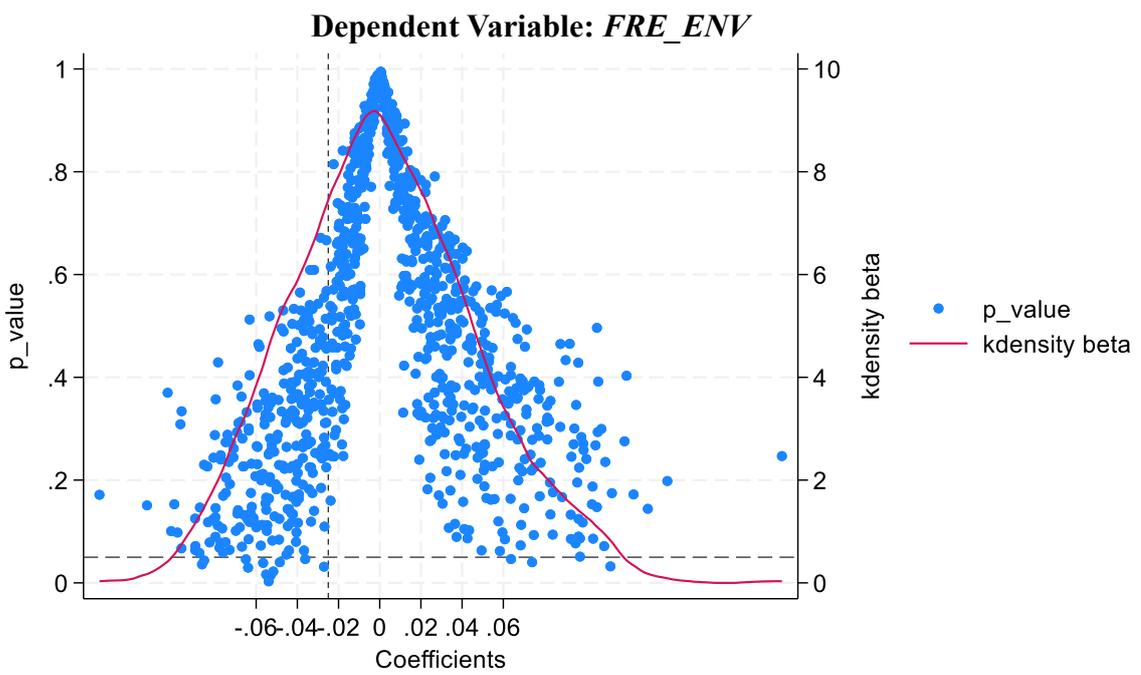

**Figure 2. Placebo test (distribution of estimated coefficients after randomization)**

**Table 1. Summary Statistics**

| Variable | Obs | Mean | SD | Min | P25 | Median | P75 | Max |
|---|---|---|---|---|---|---|---|---|
| *DUM_ENV* | 6718 | 0.048 | 0.215 | 0.000 | 0.000 | 0.000 | 0.000 | 1.000 |
| *FRE_ENV* | 6718 | 0.054 | 0.276 | 0.000 | 0.000 | 0.000 | 0.000 | 4.127 |
| *DID* | 6718 | 0.172 | 0.377 | 0.000 | 0.000 | 0.000 | 0.000 | 1.000 |
| *Age* | 6718 | 2.897 | 0.311 | 1.386 | 2.708 | 2.944 | 3.091 | 3.807 |
| *Size* | 6718 | 22.383 | 1.356 | 19.473 | 21.417 | 22.187 | 23.169 | 27.142 |
| *Lev* | 6718 | 0.405 | 0.206 | 0.050 | 0.237 | 0.391 | 0.552 | 0.979 |
| *ROA* | 6718 | 0.048 | 0.063 | -0.356 | 0.017 | 0.044 | 0.079 | 0.210 |
| *Cashflow* | 6718 | 19.530 | 1.654 | 14.929 | 18.447 | 19.415 | 20.550 | 24.339 |
| *DUAL* | 6718 | 0.273 | 0.446 | 0.000 | 0.000 | 0.000 | 1.000 | 1.000 |
| *Boardsize* | 6718 | 2.141 | 0.196 | 1.609 | 1.946 | 2.197 | 2.197 | 2.708 |
| *Top1* | 6718 | 34.932 | 14.784 | 8.340 | 23.590 | 32.755 | 44.630 | 74.950 |
| *IndRatio* | 6718 | 37.178 | 5.062 | 33.330 | 33.330 | 33.330 | 42.860 | 57.140 |

**Table 2. Baseline Regression Results**

| Variable | (1) DUM_ENV | (2) DUM_ENV | (3) FRE_ENV | (4) FRE_ENV |
|---|---|---|---|---|
| DID | -0.0295** | -0.0281** | -0.0379** | -0.0365** |
|  | (-2.644) | (-2.466) | (-2.829) | (-2.723) |
| Age |  | 0.0465 |  | 0.0671 |
|  |  | (0.710) |  | (0.818) |
| Size |  | -0.0025 |  | 0.0021 |
|  |  | (-0.405) |  | (0.327) |
| Lev |  | -0.0003 |  | -0.0130 |
|  |  | (-0.008) |  | (-0.345) |
| ROA |  | -0.0459 |  | -0.0283 |
|  |  | (-0.757) |  | (-0.402) |
| Cashflow |  | 0.0004 |  | 0.0001 |
|  |  | (0.073) |  | (0.024) |
| DUAL |  | 0.0030 |  | 0.0015 |
|  |  | (0.522) |  | (0.221) |
| Boardsize |  | 0.0085 |  | 0.0086 |
|  |  | (0.276) |  | (0.169) |
| Top1 |  | 0.0001 |  | 0.0004 |
|  |  | (0.246) |  | (0.664) |
| IndRatio |  | 0.0000 |  | -0.0012 |
|  |  | (0.045) |  | (-0.882) |
| Constant | 0.0535*** | -0.0555 | 0.0601*** | -0.1652 |
|  | (26.237) | (-0.147) | (25.203) | (-0.419) |
| Firm FE | YES | YES | YES | YES |
| Year FE | YES | YES | YES | YES |
| N | 6,718 | 6,718 | 6,718 | 6,718 |
| Adj. $R^2$ | 0.184 | 0.183 | 0.282 | 0.281 |

Notes: This table shows the regression coefficients for testing the impact of ASIOI on corporate environmental violations. The dependent variables are **DUM_ENV** and **FRE_ENV**. Year and Firm fixed effects are controlled. All other variables are defined in Appendix 1. T-statistics, reported in parentheses, are based on standard errors clustered by industry level. Significance levels of 10%, 5% and 1% are indicated by *, ** and ***, respectively.

**Table 3. Parallel trend test**

| Variable | (1) DUM_ENV | (2) DUM_ENV | (3) FRE_ENV | (4) FRE_ENV |
|---|---|---|---|---|
| $Before^{-3}$ | -0.0357 | -0.0367 | -0.0242 | -0.0250 |
|  | (-1.432) | (-1.420) | (-0.986) | (-0.995) |
| $Before^{-2}$ | -0.0273 | -0.0274 | -0.0100 | -0.0098 |
|  | (-1.430) | (-1.453) | (-0.411) | (-0.404) |
| Current | -0.0395 | -0.0389 | -0.0313 | -0.0307 |
|  | (-1.474) | (-1.452) | (-1.245) | (-1.208) |
| $After^{1}$ | -0.0473** | -0.0462** | -0.0420 | -0.0413 |
|  | (-2.388) | (-2.285) | (-1.516) | (-1.440) |
| $After^{2}$ | -0.0529** | -0.0517** | -0.0583** | -0.0582** |
|  | (-2.534) | (-2.432) | (-2.274) | (-2.220) |
| $After^{3}$ | -0.0778** | -0.0763** | -0.0899* | -0.0895* |
|  | (-2.868) | (-2.762) | (-1.951) | (-1.943) |
| Age |  | 0.0529 |  | 0.0664 |
|  |  | (0.705) |  | (0.759) |
| Size |  | -0.0020 |  | 0.0019 |
|  |  | (-0.326) |  | (0.313) |
| Lev |  | -0.0039 |  | -0.0202 |
|  |  | (-0.105) |  | (-0.599) |
| ROA |  | -0.0455 |  | -0.0282 |
|  |  | (-0.729) |  | (-0.397) |
| Cashflow |  | 0.0002 |  | -0.0000 |
|  |  | (0.031) |  | (-0.006) |
| DUAL |  | 0.0026 |  | 0.0016 |
|  |  | (0.447) |  | (0.228) |
| Boardsize |  | 0.0101 |  | 0.0103 |
|  |  | (0.313) |  | (0.195) |
| Top1 |  | 0.0002 |  | 0.0004 |
|  |  | (0.300) |  | (0.824) |
| IndRatio |  | 0.0001 |  | -0.0012 |
|  |  | (0.087) |  | (-0.851) |
| Constant | 0.0637*** | -0.0748 | 0.0668*** | -0.1529 |
|  | (9.487) | (-0.179) | (9.836) | (-0.353) |
| Firm FE | Yes | Yes | Yes | Yes |
| Year FE | Yes | Yes | Yes | Yes |
| N | 6,718 | 6,718 | 6,718 | 6,718 |
| Adj. $R^2$ | 0.185 | 0.184 | 0.283 | 0.282 |

Notes: This table presents the parallel trend estimation results. Year and Firm fixed effects are controlled. All other variables are defined in Appendix 1. T-statistics, reported in parentheses, are based on standard errors clustered by industry level. Significance levels of 10%, 5% and 1% are indicated by *, ** and ***, respectively.

**Table 4. Change the Regression Model**

| Variable | (1) DUM_ENV | (2) DUM_ENV |
|---|---|---|
| DID | -0.8673*** | -0.9264*** |
|  | (-2.804) | (-2.849) |
| Age |  | 0.7146 |
|  |  | (0.326) |
| Size |  | -0.2132 |
|  |  | (-0.576) |
| Lev |  | -0.3143 |
|  |  | (-0.291) |
| ROA |  | -1.0274 |
|  |  | (-0.496) |
| Cashflow |  | 0.0117 |
|  |  | (0.117) |
| DUAL |  | 0.1002 |
|  |  | (0.332) |
| Boardsize |  | -0.2593 |
|  |  | (-0.249) |
| Top1 |  | 0.0191 |
|  |  | (1.068) |
| IndRatio |  | 0.0078 |
|  |  | (0.228) |
| _cons | -18.2383 | -16.8253 |
|  | (-0.016) | (-0.015) |
| Firm FE | YES | YES |
| Year FE | YES | YES |
| N | 1,191 | 1,191 |
| Pseudo $R^2$ | 0.1991 | 0.2006 |

Notes: This table presents the results of logit regression on the **DUM_ENV** variable The control variables are controlled. All other variables are defined in Appendix 1. T-statistics, reported in parentheses. Year and Firm fixed effects are controlled. Significance levels of 10%, 5% and 1% are indicated by *, ** and ***, respectively.

**Table 5.CSDID**

| Variable | (1) DUM_ENV | (2) FRE_ENV |
|---|---|---|
| ASIOI | -0.1245*** | -0.0972*** |
|  | (-4.068) | (-2.861) |
| Controls | YES | YES |
| Firm FE | YES | YES |
| Year FE | YES | YES |
| N | 1,921 | 1,921 |

Notes: This table presents the results of heterogeneous treatment effects at multiple points in time. The control variables are controlled. All other variables are defined in Appendix 1. T-statistics, reported in parentheses. Year and Firm fixed effects are controlled. Significance levels of 10%, 5% and 1% are indicated by *, ** and ***, respectively.

**Table 6. Changing the clustering level**

| Variable | (1) DUM_ENV | (2) DUM_ENV | (3) FRE_ENV | (4) FRE_ENV |
|---|---|---|---|---|
| DID | -0.0295** | -0.0281** | -0.0379*** | -0.0365** |
|  | (-2.579) | (-2.454) | (-2.663) | (-2.538) |
| Age |  | 0.0465 |  | 0.0671 |
|  |  | (0.756) |  | (0.975) |
| Size |  | -0.0025 |  | 0.0021 |
|  |  | (-0.364) |  | (0.253) |
| Lev |  | -0.0003 |  | -0.0130 |
|  |  | (-0.008) |  | (-0.314) |
| ROA |  | -0.0459 |  | -0.0283 |
|  |  | (-0.818) |  | (-0.354) |
| Cashflow |  | 0.0004 |  | 0.0001 |
|  |  | (0.102) |  | (0.032) |
| DUAL |  | 0.0030 |  | 0.0015 |
|  |  | (0.329) |  | (0.155) |
| Boardsize |  | 0.0085 |  | 0.0086 |
|  |  | (0.222) |  | (0.159) |
| Top1 |  | 0.0001 |  | 0.0004 |
|  |  | (0.314) |  | (0.653) |
| IndRatio |  | 0.0000 |  | -0.0012 |
|  |  | (0.040) |  | (-0.625) |
| Constant | 0.0535*** | -0.0555 | 0.0601*** | -0.1652 |
|  | (27.113) | (-0.197) | (24.552) | (-0.507) |
| Firm FE | YES | YES | YES | YES |
| Year FE | YES | YES | YES | YES |
| N | 6,718 | 6,718 | 6,718 | 6,718 |
| Adj. $R^2$ | 0.184 | 0.183 | 0.282 | 0.281 |

Notes: This table also shows the effect of ASIOI on corporate environmental violations with the standard errors clustered at the firm level rather than industry level. Year and Firm fixed effects are controlled. All other variables are defined in Appendix 1. T-statistics, reported in parentheses, are based on standard errors clustered by industry level. Significance levels of 10%, 5% and 1% are indicated by *, ** and ***, respectively.

**Table 7. Changing the Fixed Effects**

| Variable | (1) DUM_ENV | (2) DUM_ENV | (3) FRE_ENV | (4) FRE_ENV |
|---|---|---|---|---|
| DID | -0.0320** | -0.0325** | -0.0389** | -0.0396** |
|  | (-2.424) | (-2.357) | (-2.494) | (-2.469) |
| Age |  | 0.0114 |  | 0.0288 |
|  |  | (0.166) |  | (0.349) |
| Size |  | 0.0001 |  | 0.0085 |
|  |  | (0.015) |  | (0.828) |
| Lev |  | -0.0123 |  | -0.0354 |
|  |  | (-0.345) |  | (-1.082) |
| ROA |  | -0.0524 |  | -0.0620 |
|  |  | (-0.823) |  | (-0.906) |
| Cashflow |  | 0.0003 |  | -0.0005 |
|  |  | (0.056) |  | (-0.086) |
| DUAL |  | -0.0005 |  | -0.0010 |
|  |  | (-0.104) |  | (-0.158) |
| Boardsize |  | 0.0120 |  | 0.0095 |
|  |  | (0.396) |  | (0.180) |
| Top1 |  | 0.0004 |  | 0.0007 |
|  |  | (0.680) |  | (1.288) |
| IndRatio |  | 0.0000 |  | -0.0013 |
|  |  | (0.042) |  | (-0.864) |
| Constant | 0.0539*** | -0.0205 | 0.0603*** | -0.1823 |
|  | (23.691) | (-0.050) | (22.449) | (-0.467) |
| Firm FE | YES | YES | YES | YES |
| Year_Industry FE | YES | YES | YES | YES |
| N | 6,718 | 6,718 | 6,718 | 6,718 |
| Adj. $R^2$ | 0.193 | 0.192 | 0.301 | 0.300 |

Notes: This table is the result of replacement fixed effects, controlling for firm, year-industry fixed effects. All other variables are defined in Appendix 1. T-statistics, reported in parentheses, are based on standard errors clustered by industry level. Significance levels of 10%, 5% and 1% are indicated by *, ** and ***, respectively.

**Table 8. Mechanism test: Internal control**

| Variable | (1) IC_Index | (2) IC_Index |
|---|---|---|
| DID | 0.1500* | 0.1153* |
|  | (1.974) | (1.840) |
| Age |  | 0.6775* |
|  |  | (1.873) |
| Size |  | 0.4765*** |
|  |  | (5.162) |
| Lev |  | -1.1544*** |
|  |  | (-5.739) |
| ROA |  | 2.5072** |
|  |  | (2.617) |
| Cashflow |  | -0.0078 |
|  |  | (-0.363) |
| DUAL |  | 0.1752* |
|  |  | (1.773) |
| Boardsize |  | 0.2041 |
|  |  | (0.601) |
| Top1 |  | 0.0007 |
|  |  | (0.195) |
| IndRatio |  | 0.0104 |
|  |  | (1.124) |
| Constant | 6.1162*** | -6.9279** |
|  | (478.664) | (-2.723) |
| Firm FE | Yes | Yes |
| Year FE | Yes | Yes |
| N | 6,220 | 6,220 |
| Adj. $R^2$ | 0.331 | 0.357 |

Notes: This table presents the mechanism test results. China Dib internal Control Index is used to measure internal control (*IC_Index*). A higher value of this indicator suggests stronger internal control within the firm. Year and Firm fixed effects are controlled. All other variables are defined in Appendix 1. T-statistics, reported in parentheses, are based on standard errors clustered by industry level. Significance levels of 10%, 5% and 1% are indicated by *, ** and ***, respectively.

**Table 9. Mechanism test: Green investment**

| Variable | (1) GInvest | (2) GInvest |
|---|---|---|
| DID | 0.9320* | 1.2205*** |
|  | (2.096) | (4.525) |
| Age |  | -2.7654 |
|  |  | (-0.442) |
| Size |  | 2.5155** |
|  |  | (2.240) |
| Lev |  | -0.9762 |
|  |  | (-0.344) |
| ROA |  | 5.3745 |
|  |  | (0.823) |
| Cashflow |  | -0.0848 |
|  |  | (-0.520) |
| DUAL |  | 2.2400** |
|  |  | (2.708) |
| Boardsize |  | -3.0615 |
|  |  | (-0.559) |
| Top1 |  | -0.0108 |
|  |  | (-0.148) |
| IndRatio |  | -0.1469 |
|  |  | (-1.564) |
| Constant | 13.3170*** | -22.0720 |
|  | (125.209) | (-0.879) |
| Firm FE | YES | YES |
| Year FE | YES | YES |
| N | 1,222 | 1,222 |
| Adj. $R^2$ | 0.138 | 0.147 |

Notes: This table presents the mechanism test results. The amount of green investment is used to measure green investment (***GInvest***). Year and Firm fixed effects are controlled. All other variables are defined in Appendix 1. T-statistics, reported in parentheses, are based on standard errors clustered by industry level. Significance levels of 10%, 5% and 1% are indicated by *, ** and ***, respectively.

**Table 10. Mechanism test: Green innovation**

| Variable | (1) Innovation | (2) Innovation | (3) F_Innovation | (4) F_Innovation |
|---|---|---|---|---|
| DID | 0.1105** | 0.1011** | 0.1247*** | 0.1019** |
|  | (2.664) | (2.497) | (3.392) | (2.897) |
| Age |  | -0.3026 |  | -0.5065** |
|  |  | (-1.751) |  | (-2.367) |
| Size |  | 0.0437 |  | 0.0193 |
|  |  | (1.687) |  | (0.806) |
| Lev |  | -0.1233 |  | -0.0553 |
|  |  | (-1.243) |  | (-0.539) |
| ROA |  | 0.1984 |  | 0.7581** |
|  |  | (1.503) |  | (2.601) |
| Cashflow |  | 0.0135* |  | 0.0020 |
|  |  | (2.086) |  | (0.139) |
| DUAL |  | 0.0484* |  | 0.0656 |
|  |  | (1.887) |  | (1.634) |
| Boardsize |  | 0.0820 |  | -0.0358 |
|  |  | (0.895) |  | (-0.301) |
| Top1 |  | -0.0034** |  | -0.0026** |
|  |  | (-2.395) |  | (-2.520) |
| IndRatio |  | 0.0059 |  | 0.0042 |
|  |  | (1.515) |  | (0.823) |
| Constant | 0.3304*** | -0.2807 | 0.3561*** | 1.3284 |
|  | (53.955) | (-0.592) | (63.921) | (1.451) |
| Firm FE | Yes | Yes | Yes | Yes |
| Year FE | Yes | Yes | Yes | Yes |
| N | 6,638 | 6,638 | 4,993 | 4,993 |
| Adj. R² | 0.609 | 0.611 | 0.638 | 0.641 |

Notes: This table presents the mechanism test results. The total number of green patent applications is used to measure the level of green Innovation (***Innovation***). The total number of green patent applications lagged by one period is further used to measure the level of green innovation (***F_Innovation***). Year and Firm fixed effects are controlled. All other variables are defined in Appendix 1. T-statistics, reported in parentheses, are based on standard errors clustered by industry level. Significance levels of 10%, 5% and 1% are indicated by *, ** and ***, respectively.

**Table 11. Heterogeneity test: Shareholding ratio of the largest shareholder**

| Variable | DUM_ENV | | FRE_ENV | |
|---|---|---|---|---|
| | (1) | (2) | (3) | (4) |
| | Top1_High | Top1_Low | Top1_High | Top1_Low |
| DID | -0.0153 | -0.0529** | -0.0230 | -0.0637** |
| | (-1.180) | (-2.839) | (-1.511) | (-2.863) |
| Age | 0.1340 | 0.0183 | 0.1148 | 0.0618 |
| | (1.015) | (0.163) | (0.799) | (0.426) |
| Size | -0.0045 | -0.0005 | -0.0026 | 0.0028 |
| | (-0.421) | (-0.029) | (-0.226) | (0.166) |
| Lev | -0.0087 | -0.0253 | -0.0209 | -0.0423 |
| | (-0.177) | (-0.461) | (-0.394) | (-0.813) |
| ROA | -0.0428 | -0.0508 | -0.0462 | -0.0088 |
| | (-0.443) | (-0.682) | (-0.287) | (-0.128) |
| Cashflow | -0.0049 | 0.0024 | -0.0041 | 0.0019 |
| | (-0.624) | (0.418) | (-0.402) | (0.299) |
| DUAL | 0.0115 | -0.0061 | 0.0092 | -0.0080 |
| | (0.691) | (-0.416) | (0.491) | (-0.530) |
| Boardsize | -0.0251 | 0.0175 | -0.0532 | 0.0429 |
| | (-0.603) | (0.292) | (-1.060) | (0.404) |
| IndRatio | -0.0004 | -0.0009 | -0.0020 | -0.0016 |
| | (-0.354) | (-0.406) | (-1.227) | (-0.468) |
| Constant | -0.0627 | -0.0229 | 0.0697 | -0.2329 |
| | (-0.140) | (-0.030) | (0.215) | (-0.273) |
| Firm FE | YES | YES | YES | YES |
| Year FE | YES | YES | YES | YES |
| N | 3,290 | 3,313 | 3,290 | 3,313 |
| Adj. $R^2$ | 0.243 | 0.135 | 0.348 | 0.191 |
| P-value | 0.079* | | 0.074* | |

Notes: This table shows the results of heterogeneity in the shareholding ratio of the largest shareholder. The data on the shareholding ratio of the largest shareholder of Dong are directly obtained from China CSMAR database, and can be divided into two groups according to the annual median: the largest shareholder with high shareholding ratio and the largest shareholder with low shareholding ratio. Year and Firm fixed effects are controlled. All other variables are defined in Appendix 1. T-statistics, reported in parentheses, are based on standard errors clustered by industry level. Significance levels of 10%, 5% and 1% are indicated by *, ** and ***, respectively.

**Table 12. Heterogeneity test: Regional Judicial efficiency**

|  | DUM_ENV | | FRE_ENV | |
| --- | --- | --- | --- | --- |
|  | (1) | (2) | (3) | (4) |
| Variable | Judicial efficiency _High | Judicial Efficiency _low | Judicial efficiency _High | Judicial efficiency _low |
| DID | 0.0001 | -0.0482** | -0.0057 | -0.0584*** |
|  | (0.006) | (-2.550) | (-0.299) | (-3.163) |
| Age | 0.1109 | -0.0122 | 0.1051 | 0.0396 |
|  | (0.982) | (-0.076) | (0.949) | (0.214) |
| Size | -0.0077 | -0.0056 | -0.0115 | 0.0024 |
|  | (-0.692) | (-0.600) | (-1.133) | (0.248) |
| Lev | -0.0477 | 0.0405 | -0.0350 | 0.0024 |
|  | (-0.824) | (0.405) | (-0.494) | (0.025) |
| ROA | -0.0080 | -0.0778 | 0.1510 | -0.1426 |
|  | (-0.134) | (-0.724) | (1.407) | (-1.189) |
| Cashflow | 0.0073 | -0.0062 | 0.0083 | -0.0076 |
|  | (1.136) | (-0.974) | (1.463) | (-1.008) |
| DUAL | 0.0045 | 0.0110 | 0.0026 | 0.0092 |
|  | (0.288) | (1.109) | (0.146) | (0.634) |
| Boardsize | -0.0771 | 0.0664* | -0.0756 | 0.0691 |
|  | (-1.430) | (1.804) | (-1.115) | (1.034) |
| Top1 | -0.0005 | 0.0007 | -0.0009 | 0.0014 |
|  | (-0.648) | (0.560) | (-0.874) | (1.081) |
| IndRatio | -0.0023 | 0.0022 | -0.0026 | 0.0004 |
|  | (-1.406) | (1.327) | (-1.449) | (0.200) |
| Constant | 0.0336 | 0.0829 | 0.1315 | -0.1520 |
|  | (0.081) | (0.109) | (0.291) | (-0.173) |
| Firm FE | YES | YES | YES | YES |
| Year FE | YES | YES | YES | YES |
| N | 3,132 | 3,479 | 3,132 | 3,479 |
| Adj. $R^2$ | 0.193 | 0.208 | 0.304 | 0.295 |
| P-value | 0.031** | | 0.048** | |

Notes: This table presents the results of heterogeneity in regional judicial efficiency. Judicial efficiency is measured by the development of market intermediary organizations and the legal system environment in Fan Gang index, which is divided into two groups according to the annual median: high judicial efficiency and low judicial efficiency. Year and Firm fixed effects are controlled. All other variables are defined in Appendix 1. T-statistics, reported in parentheses, are based on standard errors clustered by industry level. Significance levels of 10%, 5% and 1% are indicated by *, ** and ***, respectively.

**Table 13. Heterogeneity test: Executive Shareholding**

|  | DUM_ENV | | FRE_ENV | |
|---|---|---|---|---|
|  | (1) | (2) | (3) | (4) |
| Variable | Executive Shareholding _High | Executive Shareholding _low | Executive Shareholding _High | Executive Shareholding _low |
| DID | -0.0022 | -0.0398** | 0.0048 | -0.0445** |
|  | (-0.139) | (-2.312) | (0.316) | (-2.806) |
| Age | 0.1307 | 0.0977 | 0.1955 | 0.0869 |
|  | (0.826) | (1.242) | (1.141) | (1.195) |
| Size | -0.0031 | -0.0035 | 0.0068 | -0.0003 |
|  | (-0.244) | (-0.408) | (0.527) | (-0.030) |
| Lev | -0.0898 | 0.0176 | -0.1069 | -0.0002 |
|  | (-1.335) | (0.298) | (-1.477) | (-0.004) |
| ROA | 0.1598* | -0.1230* | 0.0835 | -0.0433 |
|  | (1.885) | (-1.801) | (0.950) | (-0.459) |
| Cashflow | 0.0064 | -0.0007 | 0.0060 | -0.0014 |
|  | (0.547) | (-0.141) | (0.493) | (-0.183) |
| DUAL | -0.0032 | 0.0001 | 0.0033 | 0.0035 |
|  | (-0.183) | (0.006) | (0.203) | (0.255) |
| Boardsize | -0.0209 | 0.0528 | -0.0191 | 0.0814 |
|  | (-0.466) | (1.563) | (-0.471) | (1.286) |
| Top1 | -0.0005 | 0.0003 | -0.0003 | 0.0007 |
|  | (-0.737) | (0.283) | (-0.371) | (0.759) |
| IndRatio | 0.0008 | -0.0004 | 0.0005 | -0.0021 |
|  | (0.451) | (-0.266) | (0.256) | (-0.859) |
| Constant | -0.3291 | -0.2409 | -0.7144 | -0.2695 |
|  | (-0.437) | (-0.536) | (-0.973) | (-0.558) |
| Firm FE | YES | YES | YES | YES |
| Year FE | YES | YES | YES | YES |
| N | 2,590 | 3,637 | 2,590 | 3,637 |
| Adj. $R^2$ | 0.062 | 0.239 | 0.067 | 0.318 |
| P-value | 0.086* | | 0.084* | |

Notes: This table presents the results based on the heterogeneity of executive shareholding. The executive shareholding ratio is measured using the chairman's shareholding ratio, which is directly obtained from the CSMAR database. Firms are divided into two groups—high and low executive shareholding—based on the annual industry median. Year and Firm fixed effects are controlled. All other variables are defined in Appendix 1. T-statistics, reported in parentheses, are based on standard errors clustered by industry level. Significance levels of 10%, 5% and 1% are indicated by *, ** and ***, respectively.

**Table 14. Heterogeneity test: ESG score**

| Variable | DUM_ENV | | FRE_ENV | |
|---|---|---|---|---|
| | (1) | (2) | (3) | (4) |
| | ESG_High | ESG_low | ESG_High | ESG_low |
| DID | -0.0067 | -0.0422** | -0.0058 | -0.0519** |
| | (-0.487) | (-2.526) | (-0.281) | (-2.633) |
| Age | -0.0432 | 0.0326 | 0.0122 | 0.0517 |
| | (-0.508) | (0.293) | (0.101) | (0.491) |
| Size | -0.0003 | -0.0053 | -0.0022 | -0.0037 |
| | (-0.038) | (-0.606) | (-0.163) | (-0.335) |
| Lev | -0.0033 | -0.0087 | 0.0031 | 0.0088 |
| | (-0.081) | (-0.188) | (0.064) | (0.122) |
| ROA | -0.2240* | 0.0593 | -0.1850 | 0.0853 |
| | (-1.974) | (0.704) | (-1.566) | (0.762) |
| Cashflow | 0.0140*** | -0.0091 | 0.0123** | -0.0054 |
| | (3.405) | (-1.156) | (2.238) | (-0.559) |
| DUAL | -0.0130** | 0.0070 | -0.0178*** | 0.0057 |
| | (-2.363) | (0.531) | (-3.671) | (0.402) |
| Boardsize | 0.0431 | -0.0178 | 0.0728 | -0.0317 |
| | (1.152) | (-0.412) | (1.154) | (-0.444) |
| Top1 | 0.0002 | -0.0001 | 0.0006 | 0.0005 |
| | (0.210) | (-0.144) | (0.581) | (0.460) |
| IndRatio | 0.0003 | -0.0002 | -0.0001 | -0.0016 |
| | (0.189) | (-0.185) | (-0.058) | (-1.625) |
| Constant | -0.1784 | 0.2987 | -0.3200 | 0.1909 |
| | (-0.400) | (0.548) | (-0.552) | (0.448) |
| Firm FE | YES | YES | YES | YES |
| Year FE | YES | YES | YES | YES |
| N | 3,189 | 3,178 | 3,189 | 3,178 |
| Adj. $R^2$ | 0.189 | 0.234 | 0.329 | 0.287 |
| P-value | 0.075* | | 0.085* | |

Notes: This table presents the results of ESG score heterogeneity. The ESG score index is directly obtained from the CNRD database of China, and is divided into two groups according to the median of the year: high ESG score group and low ESG score group. Year and Firm fixed effects are controlled. All other variables are defined in Appendix 1. T-statistics, reported in parentheses, are based on standard errors clustered by industry level. Significance levels of 10%, 5% and 1% are indicated by *, ** and ***, respectively.

**Appendix 1. Variable Definitions**

| Variables | Definition |
|---|---|
| DUM_ENV | If an enterprise is subject to environmental punishment in the current year, it is 1; otherwise, it is 0 |
| FRE_ENV | The natural logarithm of 1 plus the number of environmental violations in the current year |
| DID | For local SOEs, it is equal to 1 if ASIOI is implemented in the location of the SOEs in a given year, and 0 otherwise. For central SOEs, the value is 1 if the year is after 2018 or later, and 0 otherwise |
| IC_Index | Dib Internal Control Index |
| GInvest | The logarithm of the company's total green investment in the current year plus one |
| Innovation | Logarithm of the annual number of green patent applications of the firm plus one |
| Age | The number of years the company has been listed |
| Size | Natural logarithm of the total assets |
| Lev | Ratio of total liabilities to total assets |
| ROA | The ratio of net profit to total assets |
| Cashflow | Natural logarithm of net cash flows from operating activities |
| DUAL | Whether the CEO of the company concurrently serves as the chairman of the board of directors |
| Boardsize | Natural logarithm of the size of the firm's board of directors |
| Top1 | Shareholding ratio of the company's largest shareholder |
| IndRatio | Proportion of independent directors in the company |
| Executive Shareholding | Shareholding ratio of the chairman of the company |
| ESG | ESG score of enterprises directly disclosed by CNRD database |
| Judicial efficiency | Data on the development of market intermediary organizations and legal system environment in Fan Gang index |